\documentclass[]{piparticle-final}
\usepackage{graphicx}
\usepackage{amsmath}

\usepackage{cite} % To compress citation ranges
\usepackage{epstopdf}

\begin{document}
\volume{5}               % To be inserted by Editor
\articlenumber{050006}   % To be inserted by Editor
\journalyear{2013}       % To be inserted by Editor
\editor{G. B. Mindlin}   % To be inserted by Editor
%\reviewers{Reviewer's name}  % To be inserted by Editor
\received{14 June 2013}     % To be inserted by Editor
\accepted{9 July 2013}   % To be inserted by Editor
\runningauthor{A. Zylberberg \itshape{et al.}}  % To be inserted by Editor
\doi{050006}         % To be inserted by Editor

\title{A neuronal device for the control of multi-step computations}

% Institution references with \cite are inserted after \maketitle in theaffiliation enviroment
\author{Ariel Zylberberg,\cite{equal,lni,liaa,viscog}\thanks{E-mail: arielz@df.uba.ar}\hspace{0.5em} Luciano Paz,\cite{equal,lni}\thanks{E-mail: lpaz@df.uba.ar}\hspace{0.5em} Pieter R. Roelfsema,\cite{viscog,intne,psydep}\thanks{E-mail: p.roelfsema@nin.knaw.nl}\hspace{0.5em} Stanislas Dehaene,\cite{collfr,inserm,cea,paris}\thanks{E-mail: stanislas.dehaene@gmail.com}\hspace{0.5em} Mariano Sigman\cite{lni}\thanks{E-mail: sigman@df.uba.ar}}

\pipabstract{
We describe the operation of a neuronal device which embodies the computational principles of the ``paper-and-pencil'' machine envisioned by Alan Turing. 
The network is based on principles of cortical organization. We develop a plausible solution to implement pointers and investigate
how neuronal circuits may instantiate the basic operations involved in assigning a value to a variable ({\it i.e.}, $x=5$), in determining whether two variables have the same value 
and in retrieving the value of a given variable to be accessible to other nodes of the network. 
We exemplify the collective function of the network in simplified arithmetic and problem solving (blocks-world) tasks.

}

\maketitle

\blfootnote{
\begin{theaffiliation}{99}
   \institution{equal} These authors contributed equally to this work
   \institution{lni} Laboratory of Integrative Neuroscience, Physics Department, FCEyN UBA and IFIBA, CONICET; Pabell\'on 1, Ciudad Universitaria, 1428 Buenos Aires, Argentina.
   \institution{liaa}  Laboratory of Applied Artificial Intelligence, Computer Science Department, FCEyN UBA; Pabell\'on 1, Ciudad Universitaria, 1428 Buenos Aires, Argentina.
   \institution{viscog} Department of Vision and Cognition, Netherlands Institute for Neuroscience, an institute of the Royal Netherlands Academy of Arts and Sciences, Meibergdreef 47, 1105 BA Amsterdam, The Netherlands.
   \institution{intne} Department of Integrative Neurophysiology, Center for Neurogenomics and Cognitive Research, VU University, Amsterdam, The Netherlands.
   \institution{psydep} Psychiatry Department, Academic Medical Center, Amsterdam, The Netherlands.
  \institution{collfr} Collège de France, F-75005 Paris, France.
  \institution{inserm} INSERM, U992, Cognitive Neuroimaging Unit, F-91191 Gif/Yvette, France.
  \institution{cea} CEA, DSV/I2BM, NeuroSpin Center, F-91191 Gif/Yvette, France.
  \institution{paris} University Paris-Sud, Cognitive Neuroimaging Unit, F-91191 Gif/Yvette, France.
  
\end{theaffiliation}
}

\section{Introduction}

Consider the task of finding a route in a map. You are likely to start searching the initial and final destinations, identifying possible routes between them, and then  selecting the one you  think is shorter or more appropriate. This simple example highlights how almost any task we perform is organized in a sequence of processes involving operations which we identify as ``atomic'' (here search, memory and decision-making). In contrast with the thorough knowledge of the neurophysiology underlying these atomic operations  \cite{platt1999,schall2001,Gold2007}, neuroscience is only starting to shed light on how they organize into programs \cite{roelfsema2003,romo2003flutter,moro2010neuronal}. Partly due to the difficulty of implementing compound tasks in animal models, sequential decision-making has mostly been addressed by the domains of artificial intelligence, cognitive science and psychology \cite{newell1990,anderson1998}.

Our goal is to go beyond the available neurophysiological data to show how the brain might  sequentialize operations to conform multi-step cognitive programs. We suppose the existence of elementary operations akin to Ullman's \cite{ullman1984} routines, although not limited to the visual domain. Of special relevance to our report is the body of work that has grown out of the seminal work of John Anderson \cite{anderson1998}. Building on the notion of ``production systems'' \cite{Newell1973}, Anderson and colleagues developed ACT-R as a framework for human cognition \cite{anderson1998}. It consists  of a general mechanism for selecting productions fueled by sensory, motor, goal and memory modules. The ACT-R framework emphasizes the chained nature of human cognition: at any moment in the execution of a task, information placed in buffers acts as data for the central production system, which feeds-back to these same modules. 

Despite vast recent progress in our understanding of decision formation in simple perceptual tasks \cite{Gold2007}, it  remains unresolved how the operations required by cognitive architectures may be implemented at the level of single neurons. We address some of the challenges posed by the translation of cognitive architectures to neurons: how neuronal circuits might implement a single operation, how multiple operations are arranged in a sequence, how the output of one operation is made available to the next one.

\section{Fundamental assumptions and neuronal implementation}

\subsection{The basis for single operations}

\begin{figure*}%[th]
 \begin{center}
  \includegraphics[width=0.9\textwidth]{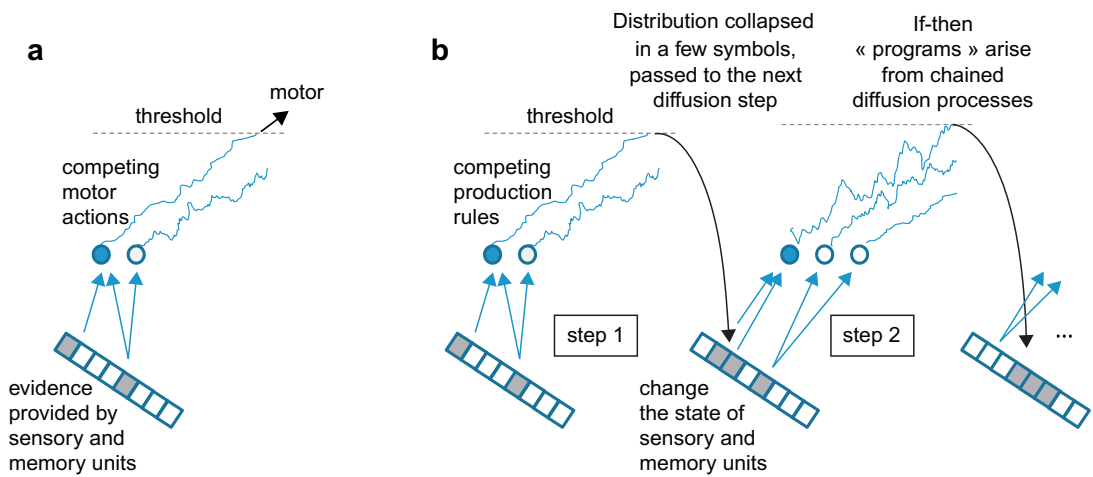}
 \end{center}
 \vspace{-20pt}
 \caption{
 Diffusion processes in single and multi-step tasks. (a) In simple sensory-motor tasks, response selection is mediated by the the accumulation of sensory evidence to a response threshold. (b) In tasks involving multiple steps, there is a parallel competition between a subset of ``productions rules''  implemented by pools of neurons that accumulate evidence up to a threshold. The selected production can have overt effects (motor actions) as well as covert effects in modifying the state of the memory, after which a new cycle begins.
 Adapted with permission from Ref. \cite{dehaene2012single}.
 } \label{fig:accumulators}
\end{figure*}

Insights into the machinery for simple sensory-motor associations come from studies of monkey electrophysiology. Studies of oculomotor decisions ---focused primarily on area LIP \cite{gottlieb2010attention}--- have shown that neurons in this area reflect the accumulation of evidence leading to a decision \cite{Gold2007}. In a well studied paradigm, monkeys were trained to discriminate the direction of motion of a patch of moving dots and report the direction of motion with an eye-movement response to the target located in the direction of motion \cite{roitman2002}. Neurons in LIP which respond with high levels of activity during memory saccades to specific portions of space are recorded during the task. These neurons show ramping activity of their firing rates with a slope that depends on the difficulty of the task, controlled by the proportion of dots that move coherently in one direction \cite{shadlen1996motion}. In a reaction time version of the task \cite{roitman2002}, when monkeys are free 
to make a 
saccade any time after the onset of the motion stimuli, the ramping activity continues until a fixed level of activity is reached, signaling the impending saccade (Fig. \ref{fig:accumulators} a). Crucially, the level of  this ``threshold'' does not depend on the difficulty of the task or the time to respond. The emerging picture is that ramping neurons in LIP integrate sensory evidence and trigger a response when activity reaches a threshold. This finding provided strong support for accumulation or race models of decision making which have been previously postulated to explain error rates and reaction times in simple tasks, and match nicely with decision theoretical notions of optimality \cite{huang2012prior}. While experimental studies have mainly characterized the feedforward flow of information from sensory to motor areas, evidence accumulation is also modulated by contextual and task-related information including prior knowledge about the likelihood and payoff of each alternative \cite{
sugrue2004matching,
wallis2001single}. Interestingly, a common currency ---the spiking activity of neurons in motor intention areas--- may underlie these seemingly unequal influences on decision formation. 

\subsection{Sequencing of multiple operations}

Brain circuits can integrate sensory evidence over hundreds of milliseconds. This illustrates how the brain decides based on unreliable evidence, averaging over time to clean up the noise. Yet the duration of single accumulation processes is constrained by its analog character, a problem pointed out  earlier by von Neumann in his book \textit{The computer and the Brain} \cite{von2012computer}: ``in the course of long calculations not only do errors add up but also those committed early in the calculation are amplified by the latter parts of it...''.  Modern computers avoid the problem of noise amplification employing digital instead of analog computation. We have suggested that the brain may deal with the amplification of noise by serially chaining individual integration stages, where the changes made by one ramp-to-threshold process represent the initial stage for the next one (Fig. \ref{fig:accumulators} b) \cite{Zylberberg2011,dehaene2012single}. 

Evidence for the ramp-to-threshold dynamics has been derived from tasks  in which the decision reflects a commitment to a motor plan \cite{maimon2006cognitive,Gold2007}. As others \cite{Shadlen2008}, we posit that the ramp-to-threshold process for action-selection is not restricted to motor actions, but may also be a mechanism to select internal actions like the decision of where to attend, what to store in memory, or what question to ask next. Therefore, the activation of a circuit based on sensory and mnemonic evidence is mediated by the accumulation of evidence in areas of the brain which can activate that specific circuit, and which compete with other internal and external actions. Within a single step, the computation proceeds in parallel and settles in a choice. Seriality is the consequence of the competitive selection of internal and external actions that transforms noisy and parallel steps of evidence accumulation into a sequence of discrete changes in the state of the network. 
These discrete steps clean 
up the noise and enable the logical flow of the computation. 

Following the terminology of symbolic cognitive architectures \cite{newell1990,anderson1998}, the ``ramping'' neurons which select the operation to do next are referred as ``production'' neurons. The competition between productions is driven by inputs from sensory and memory neurons and by the spontaneous activity in the production network. As in single-step decisions \cite{roitman2002}, the race between productions concludes when neurons encoding one production reach a decision threshold. The neurons which detect the crossing of a threshold by a production also mediate its effects, which is to activate or deactivate other brain circuits. The activated circuits can be motor, but are not restricted to it, producing different  effects like changes in the state of working memory (deciding what to remember),  activating and  broadcasting information which was in a ``latent'' state (like sensory traces and synaptic memories \cite{zylberberg2009,
mongillo2008}), or  activating peripheral processors capable of performing specific 
functions (like  changing the focus of attention,  segregating a figure from its background, or  tracing a curve).

\subsection{Pointers}

Versatility and flexibility are shared computational virtues of the human brain and of the Turing machine. The simple example of addition (at the root of Turing's conception) well illustrates  what sort of architecture this requires. One can picture the addition of two numbers x and y as displacing y steps from the initial position x. This simple representation of addition as a walk in the number line describes the core connection between movement in space and mathematical operations. It also describes the need of operations that use variables which temporarily bind to values in order to achieve flexibility. 

In this section we describe how neuronal circuits may instantiate the basic operations involved in assigning a value to a variable ({\it i.e.}, $x=5$), in determining whether two variables have the same value and in retrieving the value of a given variable. This is in a way a proof of concept, {\it i.e.}, a way to construct these operations with neurons. We are of course in no position to claim that this instantiation is unique (it is certainly not). However, we have tried to ground it on important principles of neurophsyiology and we believe that this construction raises important aspects and limitations which may generalize to other neuronal constructions of variable assignment, comparison and retrieval mechanisms.

Here we introduce the concept of pointers; individual or pool of neurons which can temporarily point to other circuits. When a pointer is active, it facilitates the activation of the circuit to which it is temporarily bound and which is dynamically set during the course of a computation. 

A pointer ``points'' to a cortical territory (for instance, to V1). This cortical mantle represents a space of values that a given variable may assume. The cortex is organized in spatial maps representing features (space, orientation, color, 1-D line...) and a pointer can temporarily bind to one of these possible set of values in a way that the activation of the pointer corresponds to the activation of the value and hence functions as a variable ({\it i.e.}, $x=3$). There are many proposed physiological mechanisms to  temporally bind neuronal circuits \cite{oreilly2006,shastri1993simple,hahnloser1999feedback}. A broad class of mechanisms relies on sustained reverberant  neuronal activity \cite{hahnloser1999feedback}. A different class relies on small modifications of synaptic activity, which constitute silent memories in latent connections \cite{mongillo2008}. Here we opt for the second alternative, first because it has a great metabolic advantage allowing to share many memories at very low cost, and 
more 
importantly because it separates the processes of variable assignation and variable retrieval. As we describe below in detail, in this architecture the current state of the variable is not broadcast to other areas until it is specifically queried.

\begin{figure*}%[th]
 \begin{center}
  \includegraphics[width=0.9\textwidth]{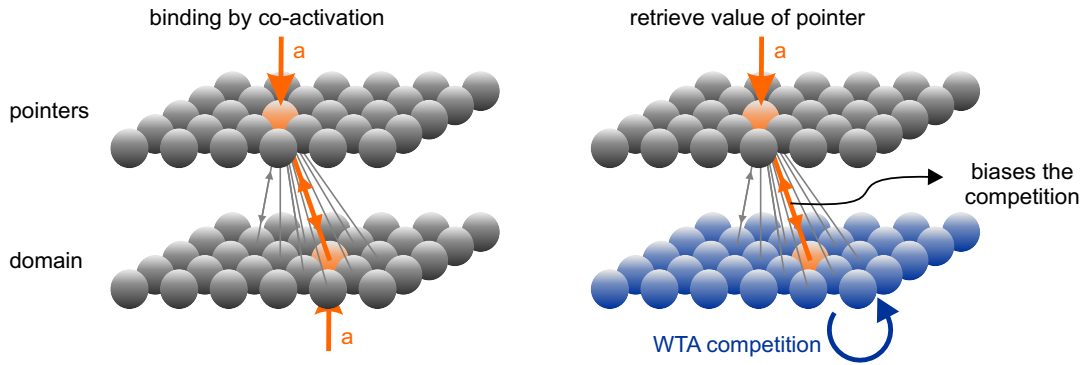}
 \end{center}
 \vspace{-20pt}
 \caption{Instantiation of variables through plastic synapses. When two neurons become active, the connection between them is rapidly strengthened, forming a transient association which decays within a few hundreds of milliseconds. The value of a pointer can be retrieved by setting the pointer's domain in a winner-take-all mode and activating the pointer which biases the WTA competition through its strengthened connections.} \label{fig:binding}
\end{figure*}

To specifically implement the binding with neuron-like elements, we follow the classic assumption that when two groups of neurons are active at the same time, a subset of the connection between them is strengthened (Fig. \ref{fig:binding}). The strengthening of the synapses is bidirectional, and it is responsible for the binding between neuronal populations. To avoid saturation of the connections and to allow for new associations to form, the strength of these connections decays exponentially within a few hundred milliseconds. Specifically, if the connection strength between a pair of populations is $w_{base}$, then when both populations are active the connection strength increases exponentially with a time constant $\tau_{rise}$ to a maximum connection strength of $w_{max}$. When one or both of the populations become inactive then the connection strength decays exponentially back to $w_{base}$ with a time constant of $\tau_{decay}$. 

The mechanism described above generates a silent coupling between a pointer and the value to which it points. How is this value recovered? In other words, how can other elements of the program know the value of the variable $X$? The expression of a value stored in silent synapses is achieved by simultaneously activating the pointer circuit and forcing the domain of the variable to a winner-take-all (WTA) mode (Fig. \ref{fig:binding}). The WTA mode ---set by having neurons with self-excitation and cross-inhibition \cite{wang2006}--- assures that only one value is retrieved at a time. These neurons make stronger connections with the neurons to which they are bound than with the other neurons. When the network is set in a WTA mode, these connections bias the competition to retrieve the value previously associated with the pointer (Fig. \ref{fig:binding}). In other words, activation of the pointer by itself is not sufficient to drive synaptic activity to the neuron (or neurons) 
representing the value to which it 
points. But it can bias the activation of a specific neuron when co-activated with a tonic activation of the entire network.

This architecture is flexible and economic. Value neurons only fire when they are set or retrieved. Memory capacity is constrained by the number of connections and not by the number of neurons. But it also has a caveat. Given that only one variable can be bound to a specific domain at any one time, multiple bindings must be addressed serially. As we show later, this can be accomplished by the sequential firing of production neurons.

\subsubsection{Compare the value of two variables}
\label{sec:compare}

If two variables $X$ and $Y$ bind to instances in the same domain, it is possible to determine whether the two variables are bound to the same instance, {\it i.e.}, whether $x=y$. The mechanics of this process is very similar to retrieving the value of a variable.

Pointer neurons\footnote{In our framework, a pointer can also be a population of neurons that functions as a single
pointer.} $X$ and $Y$ are co-activated. The equality in the assignment of $X$ and $Y$ can be identified by a coincidence detector. Specifically, this is solved by adjusting the excitability in the value domain in such a way that the simultaneous input on a single value neuron exceeds the threshold but the input of a single pointer does not. This proposed mechanism is very similar to the circuits in the brain stem which ---based on coincidence detection of delay lines--- encode interaural time  difference  \cite{carr1990circuit}. This  shows the concrete plausibility of generating such dynamic threshold mechanisms that act as coincidence detectors.

\subsubsection{Assign the value of one variable to another}

Similarly, to assign the value of $X$ to the variable $Y$ ($Y \leftarrow x$), the value of the variable that is to be copied needs to be retrieved as indicated previously, by activating the variable $X$ and forcing a WTA competition at the variable's domain. Then, the node coding for variable $Y$ must be activated, which will lead to a reinforcing of the connections between $Y$ and $x$ which will instantiate the new association. 

\section{Concrete implementation of neuronal programs}

In the previous sections we sketched a set of principles by which brain circuits can control multi-step operations and store temporary information in memory buffers to share it between different operations.

Here we demonstrate, as a proof of concept, a neuronal implementation of such circuits in two simple tasks. 

The first one is a simple arithmetic counter, where the network has to count from an initial number $n_{ini}$ to a final number $n_{end}$, a task that can be seen as the emblematic operation of a Turing device. The second example is a blocks-world problem, a paradigmatic example of multi-step decision making in artificial intelligence \cite{slaney2001blocks}. The aim of the first task is to illustrate how the different elements sketched above act in concert to implement neuronal programs. The motivation to implement blocks worlds is to link these ideas to developed notions of visual routines \cite{ullman1984,roelfsema2000,roelfsema2005}.

\subsection{Arithmetic counter}

We designed the network to be generic in the sense of being able to solve any instance of the problem, {\it i.e.}, any instantiation of $n_{ini}$ and $n_{end}$. We decided to implement a counter,  since it  constitutes  essentially a while loop and hence a basic intermediate description of most flexible computations. In the network, each node is meaningful, and all parameters were set by hand. Of course, understanding how these parameters are adjusted through a learning process is a difficult and important question, but this is left for future work. 

Each number is represented by a pool of neurons selective to the corresponding numerosity value \cite{dehaene1993development}. A potential area for the neurons belonging to the numerosity domain is the Intra Parietal Sulcus (IPS) \cite{piazza2004tuning}, where neurons coding for numerosity have been found in both humans and monkeys \cite{nieder2009representation}. In the model, number neurons interact through random lateral inhibitory connections and self-excitation. This allows, as described above, to collapse a broad distribution of number neurons \cite{dehaene1993development} to a pool representing a single number, in a retrieval process during a step of the program. 
We assume that the newtwork has learned a notion of  number proximity and continuity. This was implemented via a transition-network that has asymmetrical connections with the number-network. A given neuron representing the number $n$ excites the transition neuron $n\rightarrow n+1$ population. This in turn excites the neuron that represents the number $n+1$. Again, we do not delve into how this is learned, we assume it as a consolidated mechanism. 

The numbers-network can be in different modes: it can be quiescent, such that no number is active, or it can be in a winner-take-all mode with only one unit in the active state. Our network implementation of the counter makes use of two variables. The \textit{Count} variable stores the current count and changes dynamically as the program progresses, after being initialized to $n_{ini}$. The \textit{End} variable stores the number at which the counting has to stop and is initially set to $n_{end}$.

The network behaves basically as a while loop, increasing the value of the \textit{Count} variable while  its value differs from that of the \textit{End} variable. To increment the count, we modeled a transition-network with units that have asymmetrical connections with the numbers-network. For example, the ``$1 \rightarrow 2$'' node receives excitatory input from the unit coding for number 1 and in turn excites number 2. This network stores knowledge about successor functions, and in order to become active it requires additional input from the production system. As mentioned above, here we do not address how such structure is learned in number representing neurons. Learning to count is an effortful process during childhood \cite{lebiere1999dynamics} by which children learn transition rules between numbers. We postulate that these relations are encoded in structures which resemble horizontal connections in the cortex \cite{ts1986relationships,mcguire1991targets,gilbert1991lateral}. In 
the same way that 
horizontal connections incorporate transition probabilities of oriented elements in a slow learning process \cite{sigman2001common,gilbert2001neural}, resulting in a Gestalt as a psychological sense of ``natural'' continuity, we argue that horizontal connections between numerosity neurons can endow the same sense of transition probability and natural continuity in the space of numbers. The successor function can be as an homologous to a matrix of horizontal connections in the array of number neurons. 

In a way, our description postulates that a certain number of operations are embedded in each domain cortex (orientation selective neurons in V1 for curve tracing, number selective neurons in IPS for arithmetic...). This can be seen as ``compiled'' routines which are instantiated by local horizontal connections capable of performing operations such as collinear continuity, or ``add one''. The program can control which of these operations becomes active at any given step by gating the set of horizontal connections, a process we have referred to as ``addressing'' the cortex \cite{gilbert2007brain}. Just as an example, when older children learn to automatically count every three numbers (1, 4, 7, 10, 13...) we postulate that they have instantiated a new routine (through a slow learning process) capable of establishing the transition matrix of $n\rightarrow n+3$. The repertoire of compiled functions is dynamic and can change with learning \cite{kapadia1995improvement}.

\begin{figure*}[th]
 \begin{center}
  \includegraphics[width=0.9\textwidth]{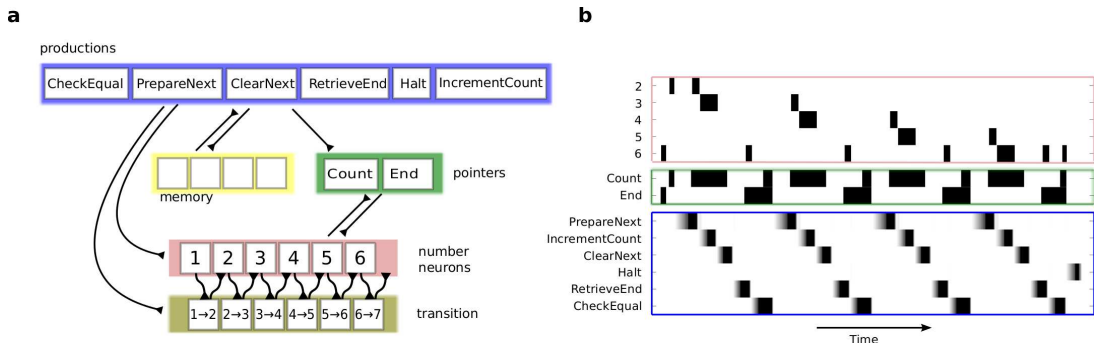}
 \end{center}
 \vspace{-20pt}
 \caption{Sketch of the network implementing an arithmetic counter. (a) The network is divided into five sub-networks: productions, memory, pointers (or variables), numbers, and transition-networks. The order in which the productions fire is controlled by the state of the memory network, which is itself controlled by the production system. (b) Dynamics of a subset of neurons in the network. All units are binary except for the production neurons (violet) which gradually accumulate evidence to a threshold.} \label{fig:counter}
\end{figure*}

Counting requires a sequence of operations which include changes in the current count, retrieval of successor functions and numeric comparisons. The successive steps of the counting routine are governed by firing of production neurons. The order in which the productions fire is controlled by the content of the memory (Fig. \ref{fig:counter}). We emphasize that while the production selection process proceeds in parallel ---as each production neuron constantly evaluates its input--- the selected production strongly inhibits the other production neurons and therefore the evidence accumulated at one step is for the most part lost after a production is selected.\footnote{In the absence of external noise (as in the present simulations), only the production with the largest input has higher-than-baseline activity.} In Fig. \ref{fig:counter} we simulate a network that has to count from numbers 2 to 6. Once the initial and final numbers have been bound to the \textit{Count} and \textit{End} 
variables respectively, the 
network cycles through six productions. The first production that is selected is the \emph{PrepareNext} production, whose role is to retrieve the value that results from adding $1$ to the current count. To this end, this production retrieves the current value of the \textit{Count} variable, and excites the neurons of the transition-network such that the node receiving an input from the retrieved value of \textit{Count} becomes active ({\it i.e.}, if 2 is active in the numbers-network, then $2 \rightarrow 3$ becomes active in the transition-network). To assure that the retrieved value is remembered for the next step (the actual change of the current count), neurons in the transition-network are endowed with recurrent excitation, and therefore these neurons remain active until explicitly inhibited. The same production also activates a node in the memory-network which excites the \emph{IncrementCount} production, which is therefore selected next. The role of the next production (\emph{IncrementCount}) is to 
actually 
update the current count, changing the binding of the \textit{Count} variable. The \emph{IncrementCount} production inhibits all neurons in the number network, to turn it to the quiescent state. Once the network is quiescent, lateral inhibition between number nodes is released and the asymmetrical inputs from the transition-network can activate the number to which it projects. At the same time, the \emph{IncrementCount} production activates the \textit{Count} variable, which is then bound to the currently active node in the numbers-network. Notice that this two-step process between the \emph{PrepareNext} and \emph{IncrementCount} productions basically re-assigns the value of the \textit{Count} variable from  its initial value $n$ to a new value $n+1$. Using a single production to replace these two (as  tried  in earlier versions of the simulations) required activating the number and transition neurons at the same time which lead to fast and uncontrolled transitions in the numbers-
network. To increase the 
current count in a controlled manner, we settled for the two-productions solution. The \emph{IncrementCount} production also activates a memory unit that biases the competition at the next stage, in favor of the \emph{ClearNext} production. This production shuts up the activity in the transition-network, strongly inhibiting  its neurons to compensate for their recurrent excitation. Shutting the activity of these neurons is required at the next step of the routine to avoid changes in the current count when the \textit{Count} variable is retrieved to be compared with the \textit{End} variable. After \emph{ClearNext}, the \emph{RetrieveEnd} production fires which retrieves the value of the \textit{End} variable to strengthen the connections between the \textit{End} variable and the value to which it is bound. This step is required since the strength of the plastic connections decays rapidly, and therefore the instantiation of the variables will be lost if not used or reactivated 
periodically. Finally, the \emph{
CheckEqual} production is selected to determine if the \textit{Count} and \textit{End} variables are equal. If both variables are equal, a node in the memory network is activated which is detected by the \emph{Halt} production to indicate that the task has been completed; otherwise, the production that is selected next is the \emph{PrepareNext} production and the production cycle is repeated. In Fig. \ref{fig:counter}b, we show the dynamics of a subset of neurons for a network that has to count from 2 to 6.

With this example we have shown how even a seemingly very easy task such as counting (which can be encoded in up to two lines in virtually any programming language) seems to require a complex set of procedures to coordinate and stabilize all computations, when they are performed by neuronal circuits with slow building of activity and temporal decay.

\subsection{A world of blocks}

A natural  extension of the numerotopic domain used in the above example is to incorporate problems  in which the actor must interact with  its environment, and sensory and motor productions ought to be coordinated.

\begin{figure}%[th]
 \begin{center}
  \includegraphics[width=0.48\textwidth]{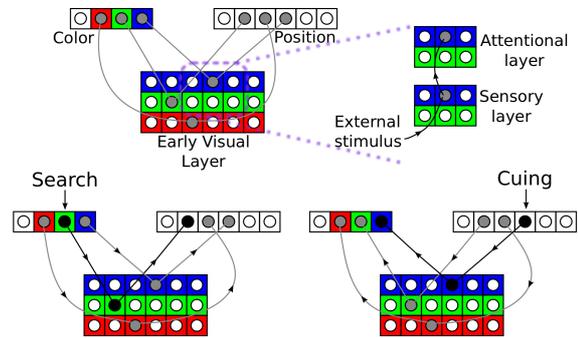}
 \end{center}
 \vspace{-20pt}
 \caption{Simplified model of the visual system used in the blocks-world simulations. The upper portion shows the different layers and their connections. The early visual area is formed by a first sensory layer of neurons that receive stimulation from the outer world and a second attentional layer with bidirectional connections between the higher color or position tuned areas. The grayed neurons are the ones that present a higher activity. The lower panels show an example of the cuing and search operations. In the  latter, a color tuned neuron is stimulated and drives an activity increase in the early visual layer. That later empowers the activity in the position layer concluding the search. The right situation shows the similar cuing operation.} \label{fig:visual}
\end{figure}

The visual system performs a great  variety of computations. It can encode a large set of visual features in a parallel feed-forward manner forming  its base representation \cite{ullman1984,lamme2000,roelfsema2000,thorpe1996}, and temporally store these features in a distributed manner \cite{felleman1991,sperling1960,graziano2008}. A matrix of lateral connections gated by top-down processing can further detect conjunctions of these feature for object recognition.

In analogy with motor routines, the visual system relies  on serial chaining of elemental operation \cite{ullman1984,roelfsema2005} to gain computational flexibility. There are many proposals as to which operations are elemental \cite{roelfsema2005}, but, as we have discussed above, this list may be fuzzy since the set of elementary operations may be changed by learning \cite{gilbert2007brain}. In this framework, atomic operations are those that are encoded in value domains.

Here, and for the purpose of implementing a neural circuit capable of solving the blocks worlds, we will focus on a simplified group of three elemental operations:
\begin{description}
 \item[Visual search:] the capacity of the system to identify the location of a given feature.
 \item[Visual cuing:] the capacity of the system of highlighting the features that are present at a given location.
 \item[Shift the processing focus:] a method guided by attention to focus the processing of visual features or other computations in a given location.
\end{description}

Here we use a simplified representation of the visual system based on previous studies \cite{hamker1999role,hamker2004dynamic,heinke2003attention}. We assume a hierarchy of two layers of neurons. The first one is tuned to conjunctions of colors and locations in the visual field. The second one has two distinct groups of populations, one with neurons that have large receptive fields that encode color irrespective of their location and another which encodes location independently of color (Fig. \ref{fig:visual}). The model assumes that neurons in the first layer tuned to a particular color and retinotopic location are reciprocally connected to second layer units that encode the same color or location.

\begin{figure}%[th]
 \begin{center}
  \includegraphics[width=0.48\textwidth]{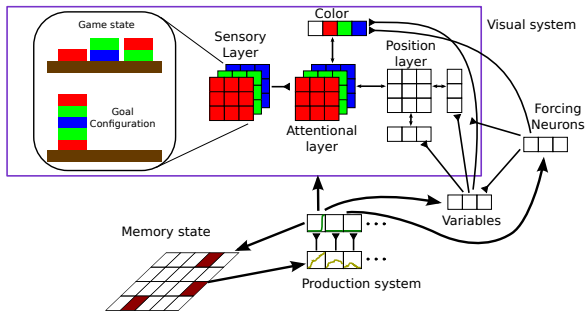}
 \end{center}
 \vspace{-20pt}
 \caption{Sketch of the network that solves the blocks-world problem. The sensory early visual system receives input from the BW configuration and excites the first attentional layer. The  latter is connected to color and position specific areas. The arrows show there is a connection between layers. The individual connections may be excitatory or inhibitory. The connections with the inverted triangle head indicate only excitatory connections exist.} \label{fig:blocks}
\end{figure}

This architecture performs visual search of color in a way which resembles the variable assignation described above, through a conjunction mechanism between maps encoding different features. In the model, the color cortex encodes each color redundantly for all positions forming a set of spatial vectors (one for each color). Of course, all these spatial maps selective for a given color can be intermingled in the cortex, as it is also the case with orientation columns which sample all orientations filling space in small receptive fields. If, for instance, a red square is presented in position three, the  neuron selective for red (henceforth referred to as in the red map) and with a receptive field in position three will fire. This activation, in turn, propagates to spatial neurons (which are insensitive to color). Thus, if four squares of different colors are presented in positions 1 to 4, the spatial neurons in these positions will fire at the same rate. To search for the spatial position of 
a red block, the 
activity of neurons coding for red in the color map must be enhanced. The enhanced activity propagates to the early visual areas which code for conjunctions of color and space, which in turn propagates to the spatial map, highlighting the position where the searched color was present. Spatial selection is triggered by an attention layer which selects the production ``attend to red'', addressing the sensory cortex in a way that only locations containing red features will be routed to the spatial neurons. The color of a block at one location can be retrieved by an almost identical mechanism. In this case, the production system sets the attentional network to a given position in space and through conjunction mechanisms (because connections are reciprocal) only the color in the selected position is retrieved. 
This is a simple device for  searching based on the propagation of attentional signals which has been used before in several models ({\it e.g.}, Ref. \cite{roelfsema2005}) (Fig. \ref{fig:visual}).

\begin{figure*}[th]
 \begin{center}
  \includegraphics[width=0.9\textwidth]{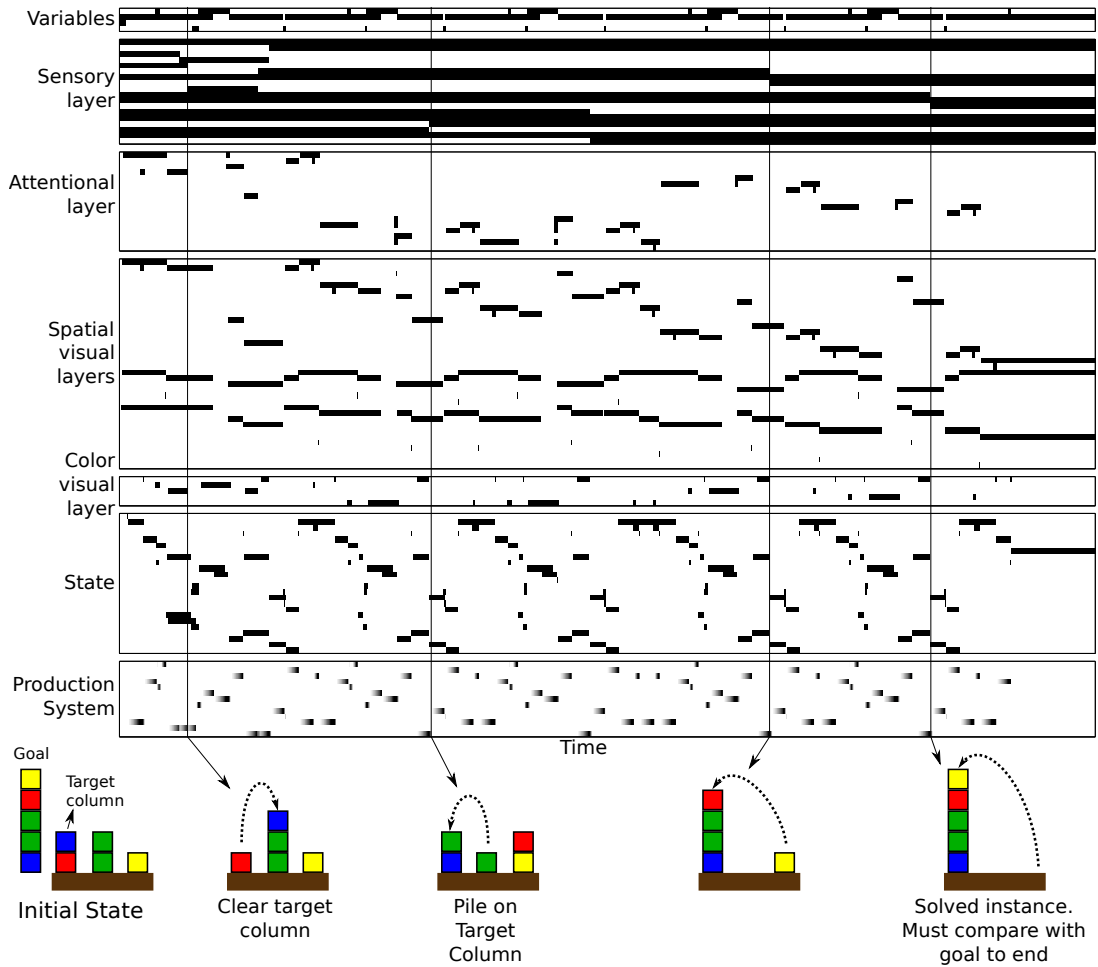}
 \end{center}
 \vspace{-20pt}
 \caption{Mean firing rate of a subset of neuronal populations involved in the resolution of an instance of BW. The rate is normalized between $0$ (white) and $1$ (black). The horizontal axis represents the ellapsed time in arbitrary units. At the bottom, we show a subset of intermediate BW configurations, aligned to the execution time of the motor commands which lead to these configurations. } \label{fig:blocks_activity}
\end{figure*}

To bridge these ideas which are well grounded in the visual literature \cite{lamme2000,roelfsema2000,roelfsema2005} with notions of planning and sequential mental programs, we use this model to implement a solver for a simple set of Blocks-World problems. The blocks-world framework is a paradigmatic artificial intelligence problem that consists of a series of colored blocks piled up on top of a large surface in many columns. The goal is to arrange the blocks according to their color in a given goal configuration. The player can only move one block from the top of any column and place it at the top of another, or on the surface that supports all the blocks. We choose to construct a solver for a restricted blocks-world problem where the surface can only hold 3 columns of blocks and the goal configuration is to arrange them all into one column (that we call the target column).\footnote{This restricted problem is equivalent to the Tower of London game \cite{shallice1982}.}

We implement a network with a set of memory and production neurons ---analogous to the counter circuit described above--- which coordinates a set of visual and motor productions (Fig. \ref{fig:blocks}). The interaction between the memory layer and the production system triggers the execution of elemental visual processes, motor actions and changes in the memory configuration in order to solve any given instance of the problem.

To solve this problem, an algorithm needs to be able to find whether a block is in the correct position. For this, it requires, first, a ``retrieve color'' from a given location function. Normally the location that is intended to be cued is the one that is being attended to. We implement the attended location as a variable population (that we call the processing focus or $PF$ inspired in Ullman's work \cite{ullman1984}), so the ``retrieve color'' is equal to cue the color in $PF$'s location.\footnote{There are works that name $PF$ as \textit{Deictic Pointers} and suggest that it would be possible to store it also by keeping gaze or even a finger at the relevant location \cite{Ballard1997}.} Second, it must compare the colors in different locations. This can be done by binding the relevant  location colors to separate variables and then comparing them in the way described in section \ref{sec:compare}

As the goal is to pile all the blocks in the correct order in a given target column, a possible first step towards the goal is to compare the target column with the goal configuration from the bottom to the top. This can be done by chaining several movements of the processing focus with color retrievals and subsequent comparisons. Once the first difference is found, the target column's upper blocks must be moved away in order to replace the different colored block with the correct one. This process is carried out using several motor productions. Once the target column is free to receive the correct colored block, that color must be searched in the remaining columns. This is done as described earlier in this section. Once found, the $PF$ can be moved there in order to view if there are blocks above it. If there are, motor productions must be chained in order to free the desired block and move it to the target column. After this is done, the program can loop back to comparing the target column with the goal 
configuration and iteratively solve the problem.

Our neuronal implementation chains the productions in a similar way as the one described above and elicits a complex activity pattern (Fig. \ref{fig:blocks_activity}). A detailed explanation of the implementation can be found in the supplementary material \cite{supplementary}.

\section{Conclusions}

Here we presented ideas aimed to bridge the gap between the neurophysiology of simple decisions and the control of sequential operations. Our framework proposes a specific set of mechanisms by which multi-step computations can be controlled by neural circuits. Action selection is determined by a parallel competition amongst competing neurons which slowly accumulate sensory and mnemonic evidence until a threshold. Actions are conceived in a broad sense, as they can result in the activation of motor circuits or other brain circuits not directly involved in a movement of the eyes or the limbs. Thresholding the action of the productions results in discrete changes to a meta-stable network. These discrete steps clean up the noise and enable a logical flow of the computation. 

Comprehending the electrophysiological mechanisms of seriality is hindered by the intrinsic difficulty of training non-human primates in complex multi-step tasks. The ideas presented in this report may serve to guide the experimental search for the mechanisms required to perform tasks involving multiple steps. Neurons integrating evidence towards a threshold should be observed even in the absence of an overt response, for instance in the frontal-eye fields of awake monkeys for the control of attention. Memory neurons should show fast transitions between metastable states, on average every $\sim 100$-$250$ msec, compatible with the mean time between successive productions in ACT-R \cite{anderson1998}. 

As mentioned, we do not address how the productions and the order in which they are executed are learned. There is a vast literature, for instance in reinforcement learning \cite{suttonbarto,roelfsema2010,Rombouts2012} describing how to learn the sequence of actions required to solve a task. Deahene \& Changeaux \cite{dehaene1997} showed how a neuronal network can solve a task similar to the BW that we modelled here, but where the order in which productions fire was controlled by the distance from the game state to the goal. Instead, our aim here was to investigate how the algorithm (the pseudo-code) may be implemented in neuronal circuits ---once it has already been learned--- from a small set of generic principles. The operation of the proposed neuronal device in a simple arithmetic task and in a neuronal network capable of solving any instance of a restricted Blocks-World domain  illustrates the plausibility of our framework for the control of computations 
involving multiple 
steps.

\begin{acknowledgements}
AZ was supported by a fellowship of the Peruilh Foundation, Faculty of Engineering, Universidad de Buenos Aires. LP was supported by a fellowship of the National Research Council of Argentina (CONICET). PRR was supported by a Netherlands Organization for Scientific Research (NWO)-VICI grant, a grant of the NWO Excellence Program Brain and Cognition, and a Human Frontier Science Program grant.
\end{acknowledgements}

%\bibliographystyle{abbrv-luis}
%\bibliography{biblio}

\end{document}